\documentclass[aps,prb,reprint,superscriptaddress]{revtex4-2}

\usepackage{graphicx}
\usepackage{dcolumn}
\usepackage{bm}
\usepackage{amsmath,amssymb,bm}
\usepackage{graphicx}
\usepackage{xspace}
\usepackage{latexsym}
\usepackage{color}
\usepackage{braket}
\usepackage[colorlinks = true,
            linkcolor = blue,
            urlcolor  = cyan,
            citecolor = cyan,
            anchorcolor = blue]{hyperref}
\usepackage{placeins}
\usepackage{orcidlink}
\usepackage{todonotes}
\usepackage{glossaries}
\usepackage{soul}
\usepackage{siunitx}
\usepackage{pythonhighlight}

\definecolor{codegreen}{rgb}{0,0.6,0}
\definecolor{codegray}{rgb}{0.5,0.5,0.5}
\definecolor{codepurple}{rgb}{0.58,0,0.82}
\definecolor{backcolour}{rgb}{0.95,0.95,0.92}

\newacronym{INS}{INS}{inelastic neutron scattering}
\newacronym{RIXS}{RIXS}{resonant inelastic X-ray scattering}
\newacronym{XAS}{XAS}{X-ray absorption spectroscopy}
\newacronym{TFY}{TFY}{total fluorescence yield}
\newacronym{QSL}{QSL}{quantum spin liquid}
\newacronym{DSF}{DSF}{dynamical structure factor}
\newacronym{1D}{1D}{one-dimensional}
\newacronym{AFM}{AFM}{antiferromagnet}
\newacronym{HAF}{HAF}{Heisenberg antiferromagnet}
\newacronym{NSC}{NSC}{non-spin-conserving}
\newacronym{SC}{SC}{spin-conserving}
\newacronym{SCO}{SCO}{Sr$_2$CuO$_3$\xspace}
\newacronym{QFI}{QFI}{quantum fisher information}
\newacronym{ED}{ED}{exact diagonalization}
\newacronym{UCL}{UCL}{ultra short core-hole lifetime}
\newacronym{NMR}{NMR}{nuclear magnetic resonance}
\newacronym{uSR}{$\mu$SR}{muon spin spectroscopy}
\newacronym{BZ}{BZ}{Brillouin zone}
\newacronym{GS}{GS}{ground state}
\newacronym{KPZ}{KPZ}{Kardar--Parisi--Zhang}
\newcommand{\SCO} {Sr$_2$CuO$_3$\xspace}
\newacronym{FWHM}{FWHM}{full-width-half-maximum}

\renewcommand{\thesection}{\arabic{section}}%

\usepackage{soul}
\usepackage{ulem}

\begin{document}

\title{Disentangling multi-spin dynamic correlations in the Heisenberg spin-$\frac{1}{2}$ chain} 

\author{V.~K.~Bhartiya\orcidlink{0000-0002-3575-7404}}
\email{vbhartiya1@bnl.gov}
\affiliation{Condensed Matter Physics and Materials Science Department, Brookhaven National Laboratory, Upton, New York 11973, USA}
\affiliation{National Synchrotron Light Source II, Brookhaven National Laboratory, Upton, New York 11973, USA}
\author{U.~Kumar\orcidlink{0000-0001-6322-9839}}
\email{kumaru@ornl.gov}
\affiliation{Materials Science and Technology Division, Oak Ridge National Laboratory, Oak Ridge, Tennessee 37831, USA}
\affiliation{Department of Physics and Astronomy, The State University of New Jersey, 136 Frelinghuysen Road, Piscataway, New Jersey 08854,  USA}
\author{T.~Kim}
\affiliation{National Synchrotron Light Source II, Brookhaven National Laboratory, Upton, New York 11973, USA}
\author{S.~Fan}
\affiliation{National Synchrotron Light Source II, Brookhaven National Laboratory, Upton, New York 11973, USA}

\author{S.~Okamoto\orcidlink{https://orcid.org/0000-0002-0493-7568}}
\affiliation{Materials Science and Technology Division, Oak Ridge National Laboratory, Oak Ridge, Tennessee 37831, USA}
\author{M. Mitrano\orcidlink{https://orcid.org/0000-0002-0102-0391}}
\affiliation{Department of Physics, Harvard University, Cambridge, Massachusetts 02138, USA}
\author{M.~P.~M.~Dean\orcidlink{0000-0001-5139-3543}}
\affiliation{Condensed Matter Physics and Materials Science Department, Brookhaven National Laboratory, Upton, New York 11973, USA}
\author{J.~Pelliciari\orcidlink{0000-0003-1508-7746}}
\affiliation{National Synchrotron Light Source II, Brookhaven National Laboratory, Upton, New York 11973, USA}

\author{I.~A.~Zaliznyak\orcidlink{0000-0002-8548-7924}}
\affiliation{Condensed Matter Physics and Materials Science Department, Brookhaven National Laboratory, Upton, New York 11973, USA}

\author{S.~Johnston\orcidlink{0000-0002-2343-0113}}
\affiliation{Department of Physics and Astronomy, The University of Tennessee, Knoxville, Tennessee 37996, USA}
\affiliation{Institute for Advanced Materials and Manufacturing, University of Tennessee, Knoxville, Tennessee 37996, USA}
\author{V.~Bisogni\orcidlink{0000-0002-7399-9930}}
\email{bisogni@bnl.gov}
\affiliation{National Synchrotron Light Source II, Brookhaven National Laboratory, Upton, New York 11973, USA}

\begin{abstract}

Higher-order correlations are essential for understanding exotic  phases and uncovering universal aspects of quantum dynamics. While inelastic neutron scattering provides well-established access to two-particle  correlations, measuring correlations in solids involving more than two particles remains a major challenge. Focusing on magnetic excitations in the Heisenberg spin-$\frac{1}{2}$ chain, we demonstrate that \gls*{RIXS} can selectively probe multi‑spin dynamical correlations by exciting distinct intermediate states through energy detuning. Through theoretical modeling, we isolate the two- and multi-spin responses, and establish that: (i) The resonant energies of the two- and multi- spin dynamical correlations are separated by $\approx \frac34 J$; implying that the multi-spin part of the cross-section comes from intermediate states that contain spin flips; (ii) The  spectral weight of the two-spin channel exhibits a Lorentzian resonant energy profile consistent with a single dominant electronic configuration, whereas the multi-spin channel response shows a Gaussian-like profile, implying  contributions from multiple intermediate spin configurations. These characteristics are reproduced by exact diagonalization of the $t-$$J$ Hamiltonian, which further reveals that the widths of Lorentzian and Gaussian resonant energy profiles depend primarily on core-hole lifetime ($\Gamma/2$) and $\frac{2J}{\Gamma}$, respectively. Controlled access to multi-spin dynamics, using \gls*{RIXS} energy detuning as a knob, can open new pathways to explore many-body dynamics in  quantum materials.

\end{abstract}

\maketitle

\glsresetall  

\section{Introduction}

Understanding exotic phases, quantum phase transitions, and universality in materials hinges on probing their elementary excitations. The ability to access the spin \gls*{DSF}---a two-point dynamic correlation function widely measured by \gls*{INS}---has revolutionized our understanding of these fields~\cite{Coldea_PRB_2003, Ruegg_PRL_2008, Coldea_Science_2010,Wu_Science_2016,Blosser_PRL_2018}. While this framework has been highly successful, the two-point correlations measured within the linear response regime often fail to fully capture the quantum many-body dynamics that encodes crucial information about the system. As a result, the limitations of this approach are now becoming apparent. The characteristic continuum of fractional spin excitations measured by \gls*{INS}, widely used to diagnose a \gls*{QSL} state, can be mimicked by chemical disorder and spin-glass behavior~\cite{Paddison_NatPhys_2017, Ma_PRL_2018, Zhu_PRL_2017}. This observational overlap suggests that reliance on two-spin dynamic correlations alone is not definitive for \gls*{QSL} state identification. Multi-particle correlations ($n$-body correlations with $n>2$) are being used increasingly in cold atom experiments and other quantum simulators to gain new insights into quantum many-body states~\cite{Bohnet_Science_2016, Schweigler_Nature_2017, Rispoli_Nature_2019, Rosenberg2024, Chalopin_arXive_2025, Chalopin_PNAS_2026}. 
For example, the once-considered universal dynamics in  Heisenberg spin-$\frac{1}{2}$ chain based on the established \gls*{KPZ} scaling of the infinite-temperature \gls*{DSF}~\cite{Ljubotina_PRL_2019, Scheie_NP_2021}, was recently brought into question by measurements on a finite-length chain of superconducting qubits~\cite{Nardis_PRL_2023,Rosenberg2024,Krajnik_PRL_2024}. These experiments demonstrated that multi-point correlations do not conform to the expected \gls*{KPZ} scaling, highlighting a regime where the two-point correlation description is inadequate and suggesting the need to revisit the material's proposed universality class. Therefore, accessing higher-order correlations that encode the system's full  dynamics is essential~\cite{Knolle_ARCMP_2019,Krajnik_PRL_2024}. While multi-particle correlations are generally accessible in quantum computers and cold atoms experiments~\cite{Bohnet_Science_2016,Schweigler_Nature_2017,Rispoli_Nature_2019,Rosenberg2024,Chalopin_arXive_2025,Chalopin_PNAS_2026}, they have generally been much harder to extract and characterize in solid state systems.

\Gls*{RIXS} has emerged  as a powerful and versatile probe of quantum materials over the past two decades~\cite{Ament_RMP_2011, Mitrano2024exploring}. Based on the resonant absorption and reemission of a photon, \gls*{RIXS} can access collective spin, charge, orbital, and lattice excitations with energy and momentum resolution~\cite{Ament_RMP_2011, Fatuzzo2015spin, LeTacon2011, MorettiSala2011, Ghiringhelli_Science_2012, Schlappa_Nature_2012, Thomas2025theory, Lu_Science_2021, Lee2021spectroscopic, Nag2024Impact}. Crucially, \gls*{RIXS} measures a response that is not fully described by linear response theory and contains contributions from both two- and multi-particle response functions~\cite{Ament2007ultrashort, Haverkort2010theory, Jia2016using}. This aspect has allowed the technique to probe novel fractionalized quasiparticles including orbitons~\cite{Schlappa_Nature_2012}, multi-spinons~\cite{Schlappa_NatComm_2018} in spin-$\frac{1}{2}$ Heisenberg chains and, quadrupolar~\cite{Nag2022quadrupolar} magnetic excitations in spin-1 Heisenberg chains. There have also been proposals to use \gls*{RIXS} to access gapless Majorana fermions in Kitaev \glspl*{QSL} \cite{Halasz_PRB_2016, Natori_PRB_2017} and measure entanglement witnesses both in- and out-of-equilibrium~\cite{ren2024witnessing, Liu2025entanglement, Hales2023witnessing}.

The primary hindrance to exploiting \gls*{RIXS}'s nonlinear capabilities is the difficulty in disentangling  the linear (two-spin) and non-linear (multi-spin) contributions to the excitation spectra. While the use of different absorption edges to emphasize specific correlation functions is well established---for example, predominantly multi-spin channels at the O $K$-edge~\cite{Schlappa_NatComm_2018} and Cu $K$-edge~\cite{Ishii2025} and two-spin channels at the Cu $L_3$-edge~\cite{PhysRevX.12.021041}---isolating these channels within a single edge remains a challenge.
At the Cu $L_3$-edge, both contributions coexist; however, their separation is obscured by overlapping resonances and complex many-body rearrangements in the intermediate state~\cite{Valentina_PRL_2014}. Achieving such resonance selectivity could provide a new route to defining \gls*{RIXS} sum rules and spectral-weight constraints, analogous to those long exploited in \gls*{INS}~\cite{Mourigal_NatPhys_2013}.
Incident energy detuning has been used to select specific intermediate states and enhance associated correlations~\cite{Nag_PRL_2020,He_NatComm_2024}. However, the role of these intermediate states has so far only been understood within the local multiplet theory framework that accurately describes \gls*{XAS} but does not account for the mobile collective excitations that are hallmarks of the quantum many-body system.

A recent report of bimagnon resonance at detuned incident energies at the Cu $L_3$-edge of La$_2$CuO$_4$~\cite{Singh_SciRep_2025} highlights that tuning to specific intermediate states provides direct access to multi-spin dynamic correlations, simultaneously capturing both two-spin (magnon) and multi-spin (bimagnon) channel contributions at a single edge. However, overlapping spectral weight from single and bimagnon excitations prevents a definite investigation of the underlying multi-spin contributions. To achieve the clearest characterization of these nonlinear effects, an ideal system must have an exact known expression for the two-particle spectral weight and/or a two-particle response that does not overlap with its multi-particle spectral weight. The \gls*{1D} \SCO cuprate satisfies both of these requirements. It is a near ideal Heisenberg spin-$\frac{1}{2}$ chain with isotropic exchange interactions and is a well-studied realization of an SU(2) symmetric ground state with fractional magnetic excitations~\cite{Fujisawa_PRB_1999, Walters_NatPhys_2009}. An exact expression for its two-particle spectral weight is provided by the Bethe ansatz~\cite{Michael_PRB_1997,Caux_JSM_2006}, and \gls*{RIXS} measurements~\cite{Schlappa_NatComm_2018} have previously observed multi-spin excitations in this system that are well separated from its two-spin excitations.

By employing Cu $L_3$-edge \gls*{RIXS} on a \SCO single crystal, we discovered that the resonant energies of two- and multi-spin dynamic correlations are well separated in energy by $\approx \frac{3}{4}J$;  where $J$ is the magnetic exchange energy. The integrated spectral weight of these dynamic correlations exhibit distinct resonant energy profiles, following Lorentzian- and Gaussian-like distributions, respectively. One key finding of this study is that these characteristics --- distinct resonant energies and resonant energy profiles --- hold true even within the two-spinon continuum, providing a robust experimental means to diagnose fractional excitations in a region where two- and multi-spin channels otherwise overlap. \Gls*{ED} calculations of the $t-$$J$ Hamiltonian quantitatively reproduce these characteristics. Our modeling shows that the observed difference in resonant energies is primarily determined by the energy cost of breaking a spin-singlet dimer into uncorrelated spins ($\approx \frac34J$), which creates parallel spin segments in the intermediate state and consequently results in a higher resonant energy. Similarly, the width of the Lorentzian and Gaussian resonant energy profiles depends largely on $\Gamma/2$ and the ratio $\frac {2J} {\Gamma}$, respectively, where $\Gamma/2$ is the inverse core-hole lifetime of the intermediate state.

\section{Extracting multi-spin dynamics}

\begin{figure*}
    \centering
    \includegraphics[width=\linewidth]{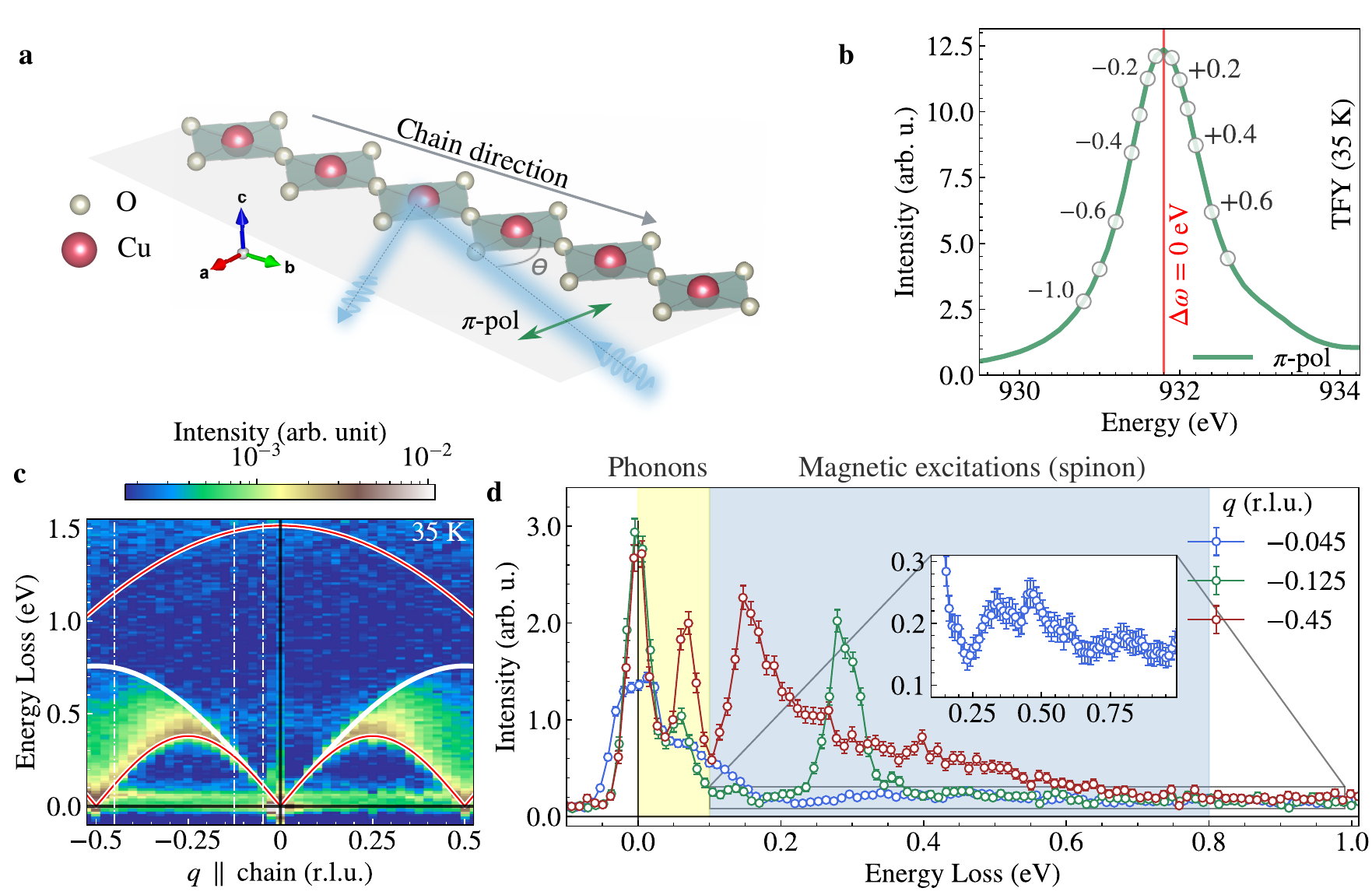}
    \caption{ (\textbf{a}) A schematic showing the $S = 1/2$ 1D spin chain in \SCO made of corner-shared Cu-O plaquettes and the experimental geometry used in this study. (\textbf{b}) \gls*{XAS} (\gls*{TFY}) measured with  $\pi$ X-ray polarization at $\theta$ = 63$^\circ$ and $\Omega$ = 150$^\circ$. The vertical red line indicates the Cu $L_3$-edge resonance, which corresponds to zero detuning $\Delta\omega = 0$. Other values of the detuning energy $\Delta\omega$ used in this study are indicated by gray circles. (\textbf{c}) A dispersion map of the magnetic excitations in \SCO with the two- and four-spinon boundaries overlaid in white and red, respectively. (\textbf{d}) Constant momentum cuts at representative $q = -0.045$, $-0.125$, and $-0.450$ (in r.l.u.) highlighting the spinon spectral weight within and outside (highlighted by an inset plot) of the two-spinon continuum.}
\label{fig:fig1}
\end{figure*}

\begin{figure*}
    \centering
     \includegraphics[width=\linewidth]{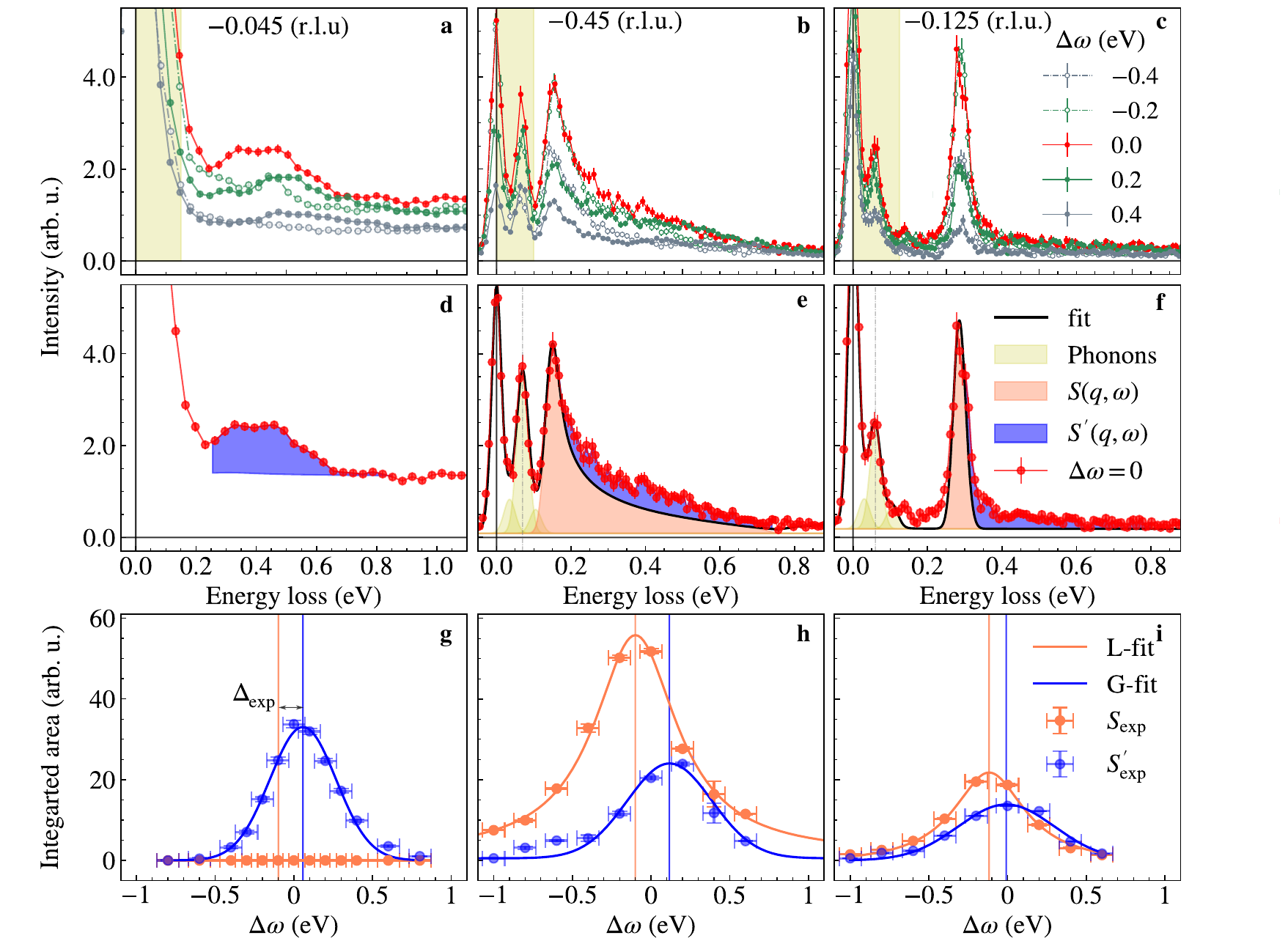}
   \caption{ (\textbf{a})-(\textbf{c}) \gls*{RIXS} spectra recorded as a function of the detuning energy $\Delta\omega$ for different momentum transfers, as indicated in each panel. 
   (\textbf{d})-(\textbf{f}) Fitted spectra taken on  resonance spectra ($\Delta\omega = 0$) shown in panels \textbf{a}-\textbf{c}, respectively. 
   The line shape labeled $S(q,\omega)$ (coral) is the exact  \gls*{DSF} for a spin-$\frac{1}{2}$ Heisenberg chain. The contribution from phonons (yellow) is accounted by the Gaussian lineshapes. The line shape labeled $S^{\prime}(q,\omega)$ (blue) is additional multi-spin spectral weight that is unique to \gls*{RIXS} measurements.  (\textbf{g})-(\textbf{i}) Integrated spectral weights  highlighting the distinct resonant energy and resonant energy profile for $S_\text{exp}$ and $S^{\prime}_\text{exp}$ (see text). ``L-fit'' and ``G-fit'' indicate Lorentzian and Gaussian fits to the data. Each horizontal row shares a common legend shown in the third column. The error bars on $\Delta\omega$ are determined from the uncertainty in determining the incident energy ($\approx$ 50 meV) and on integrated intensity ($1\sigma$) are produced from the lineshape fits. The shift in the resonance energies, denoted $\Delta_\mathrm{exp}$ panels \textbf{g}-\textbf{i} are $\Delta_\text{exp} = 0.16 \pm 0.02$, $0.20\pm 0.04$, and $0.15\pm 0.04$ for $q = -0.045$, $-0.45$, and $-0.125$ r.l.u, respectively. }
    \label{fig:fig2}
\end{figure*}

Figure~\ref{fig:fig1} presents Cu $L_3-$edge \gls*{XAS} and \gls*{RIXS} measurements on  \SCO single crystal, and demonstrates the incident energy detuning methodology for disentangling two- and multi-spin dynamical correlations. The sample was mounted such that X-ray $\pi$ ($\sigma$) polarization  lies within (perpendicular to) the CuO$_2$ plaquette, as shown in Fig.~\ref{fig:fig1}\textbf{a}. Figure~\ref{fig:fig1}\textbf{b} presents the Cu $L_3-$edge \gls*{XAS} profile measured with $\pi$-polarization. The spectrum has a single resonance centered at $\approx 931.8$ eV, matching previous results~\cite{Schlappa_Nature_2012}, and which corresponds to the excitation of a $2p_{3/2}$ core electron into the upper Hubbard band of the CuO chain. 

The spin excitations measured at the maximum of the Cu $L_{3}$-resonance are shown in Fig.~\ref{fig:fig1}\textbf{c} as a function of momentum transfer $q$. 
The spectra agree well with the 
spinon continuum observed in \gls*{INS}~\cite{Walters_NatPhys_2009} and  \gls*{RIXS}~\cite{Schlappa_Nature_2012}. Apart from the  dominant spectral weight within the boundaries of the two-spinon continuum, additional spectral weight residing outside of two-spinon phase space is visible. The spectrum also has a dominant sharp phonon excitation at $\approx$ 70 meV, which is apparent across the Brillouin zone and corresponds to the Cu-O bond stretching mode. 

Three representative high-statistics \gls*{RIXS} spectra cuts at $q = -0.045$, $-0.125$, and $-0.45$ r.l.u. in high-resolution ($\approx33$ meV) mode, are shown in Fig.~\ref{fig:fig1}\textbf{d}; these further highlight the multi-spin spectral weight together with the spectral weight within the two-spinon continuum.  The phonon and a broad spinon continuum are visible near the \gls*{BZ} boundary at $q=-0.45$ r.l.u. with similar amplitudes, while a sharp spinon peak is visible at intermediate $-0.125$ r.l.u. momentum transfer. 
Due to the weak spectral weight at higher energies in the $q = -0.045$ r.l.u. spectra [highlighted in a zoomed-in ($\times 10$) inset], we relaxed the energy resolution ($\approx 66$ meV) to gain photon flux. 
This additional spectral weight is unique to the \gls*{RIXS} cross-section and not allowed in the linear response (two-spin correlations channel) measured by \gls*{INS}~\cite{Michael_PRB_1997}. Therefore, it is suggested to have multi-spin (or multi-point) correlation character and has been attributed to excitations involving four or more spinons~\cite{Schlappa_NatComm_2018, Umesh_NewPhy_2018}. 

In the following, we demonstrate that detuning the incident X-ray energy away from the Cu $L_3$ resonance can isolate the higher-order dynamical spin correlations, enabling us to directly probe and control the relative weight of two- and multi-spin excitations inside and outside of the two-spinon continuum. Spectral profiles at selected momentum transfers  $q=-0.045$, $-0.45$, $-0.125$ r.l.u. as a function of representative detuning energies $\Delta\omega =  0$, $\pm 0.2$, $\pm 0.4$ eV, defined as the deviation from the overall $L_3$-edge resonance, are shown in Fig. \ref{fig:fig2}\textbf{a}-\textbf{c}. The yellow shading highlights the region of phonon contribution to the spectra.

The main findings of this study can be already drawn from the raw data in Figs. \ref{fig:fig2}\textbf{a} and  \ref{fig:fig2}\textbf{b}. First, for momentum transfers near the \gls*{BZ} center, the additional spectral weight peaks for incident energies close to the resonance ($\Delta \omega \approx 0$) and monotonically decreases for both positive and negative detuning. In contrast, for momentum transfer near the \gls*{BZ} boundary (Fig. \ref{fig:fig2}\textbf{b}) corresponding to broad continuum, the spectral weight near the two-spinon lower boundary resonates at lower energy compared to the spectral weight close to the two-spinon upper boundary. As a result, different sections of this broad continuum evolve differently with $\Delta \omega$. The behavior is most visible in spectrum at $\Delta \omega = 0.2$; the leading edge at 0.15 -- 0.25 eV is much lower compared to $\Delta \omega = 0$ and $-0.2$ spectra while the high-energy tails at 0.4 -- 0.7 eV are comparable across the different spectra. This behavior suggests the presence of different resonances {\it and} implies a dual character of the spin excitations within the spinon continuum. In Fig.~\ref{fig:fig2}\textbf{c}, the spinon continuum at $q = - 0.125$ r.l.u. is narrow and the peak amplitude mimics the behavior of the continuum leading edge at $q = - 0.45$ r.l.u.. Due to the narrow spinon lineshape at this $q$, the behavior of different sections of the continuum is not very evident in the raw data but is clarified with the help of additional modeling and fits.

Representative \gls*{RIXS} spectra taken at resonance ($\Delta\omega = 0$) are shown in Figs.~\ref{fig:fig2}\textbf{d}-\textbf{f}. Figures~\ref{fig:fig2}\textbf{d} show the multi-spinon contribution after removing the background determined at the maximum detuning $\Delta\omega = -0.8$ eV. The multi-spinon contribution to the spectra is already diminished at $-0.4$ eV, as seen in Fig.~\ref{fig:fig2}\textbf{a}.  Figures.~\ref{fig:fig2}\textbf{e} and \textbf{f} show fits with contributions from phonons and magnetic excitations. The background, determined at maximum detuning $\Delta\omega = -0.8$ eV, where the contribution from the multi-spin excitations are completely suppressed, was also subtracted off in this case.  


We account for the elastic and three distinct phonon contributions (35 and 70 meV modes and their combination at 105 meV~\cite{Misochko_PRB_1996}) using resolution limited Gaussian lineshapes. For the two-spin correlation function, we use an exact form for the two-spinon \gls*{DSF} given by the Bethe ansatz~\cite{Michael_PRB_1997}, which accounts for both the two- and four-spinon contributions and denoted as $S(q,\omega)$, as described in Appendix \ref{sec:DSF}. The two- and four-spinon contributions have approximately the same lineshape and nearly exhaust ($\approx 99\%$) the known sum rule for the \gls*{DSF}~\cite{Caux_JSM_2006,Walters_NatPhys_2009, Mourigal_NatPhys_2013}). Only the amplitudes associated with elastic line, phonons, and $S(q,\omega)$, were allowed to change during the fit. After accounting for these contributions, we obtain the remaining spectral weight in the \gls*{RIXS} spectra, which we denote $S^{\prime}(q,\omega)$. This contribution is specific to \gls*{RIXS} and  absent in the linear response measured in \gls*{INS} experiments. Unambiguous evidence of this is that a substantial amount of this spectral weight at $-0.045$ r.l.u (Fig.~\ref{fig:fig2}\textbf{d}), falls outside of two-spinon continuum, in contrast to $S(q,\omega)$, which vanishes in this region~\cite{Michael_PRB_1997}. The presence of a finite $S^{\prime}(q,\omega)$ within (at $q = 0.45 $ r.l.u.) and outside (at $q = 0.125 $ r.l.u.) the two-spinon boundary is more clear from the fits. Our assumption of coexisting two- and multi-spinon excitations is further justified using the \gls*{UCL} approximation $I^\text{UCL}(\boldsymbol{q},\omega)  = |W |^2 [S(\boldsymbol{q},\omega) + S^{\prime}(\boldsymbol{q},\omega)]$, where the $S(\boldsymbol{q},\omega)$ has the form of the two-spin \gls*{DSF} and  $S^{\prime}(\boldsymbol{q},\omega)$ is a more complicated multi-spin response in a Mott gapped system~\cite{Jia2016using}. \SCO is a Mott insulator with an optical gap close to 2 eV, so contributions from charge excitations are not expected at these energies. 

To extract the incident energy dependence of $S(q,\omega)$ and $S^{\prime}(q,\omega)$, Figs.~\ref{fig:fig2}\textbf{g}-\ref{fig:fig2}\textbf{i} plot the energy loss-integrated spectral weights, $S_\text{exp}(q) = \int_{\omega_\text{min}}^{\omega_\text{max}} S(q,\omega)d\omega$  and $S^{\prime}_\text{exp}(q) = \int_{\omega_\text{min}}^{\omega_\text{max}}  S^{\prime}(q,\omega)d\omega$ as a function of detuning $\Delta\omega$. $S(q,\omega)$ is nonzero only between the lower [$\omega_\mathrm{L}(q) = \frac{\pi}{2}J|\sin(2\pi q)|$] and upper [$\omega_\mathrm{U}(q) = \pi J|\sin(\pi q)|]$ boundaries of the two-spinon continuum.  The lower integration limit ($\omega_\text{min}$) for $S^\prime(q,\omega)$ was therefore set to $\omega_L(q)$ and the upper limit ($\omega_\text{max}$) was set to the maximum bandwidth of the two-spinon continuum (800 meV). The integrated spectral weights reveal two remarkable features. First, the resonant energy profiles for $S_\text{exp}(q)$ at $q=-0.45$ and $-0.125$ have Lorentzian lineshape with long tails while $S^{\prime}_\text{exp}(q)$ has a Gaussian lineshape.  We present more details on uniqueness of these fitted lineshapes in Appendix~\ref{sec:fitting_detuning}.
Crucially, the $S_\text{exp}(q)$ curves have the same resonant energy
and profiles at both momentum transfers. This behavior suggests that they represent a similar  two-spin dynamic correlations encoded in $S(q,\omega)$. Second, $S_\text{exp}(q)$ and $S^{\prime}_\text{exp}(q)$ resonate at different energies with a difference $\Delta_\text{exp}$, defined as the difference of $S_\text{exp}(q)$ and $S_\text{exp}^{\prime}(q)$ mean values, of $\approx 0.16(2)$ eV at $q  = -0.045$, 0.20(4) eV at $q = -0.45$, and 0.15(4) at $q = -0.125$. [The value of $S_\text{exp}(q= -0.45)$ is used to estimate $\Delta$ at $q = -0.045$ r.l.u., as $S_\text{exp}(q)$ vanishes near the zone center.] The distinct resonant energy of $S_\text{exp}^{\prime}(q)$ at all measured momentum transfers indicates the distinct character of the underlying spin correlations measured by $S^{\prime}(q,\omega)$. The resonant energy difference $\Delta_\text{exp}$ is  $\approx 
\frac{3}{4}J$, i.e., the energy needed to break a nearest-neighbor singlet and form two uncorrelated spins. This observation suggests that different resonances correspond to different intermediate excited states in \SCO, where the the higher energy intermediate states contain some spin excitations, costing an energy $\sim J$. 

\section{Microscopic origin}

\begin{figure*}[t]
    \centering
    \includegraphics[width=\linewidth]{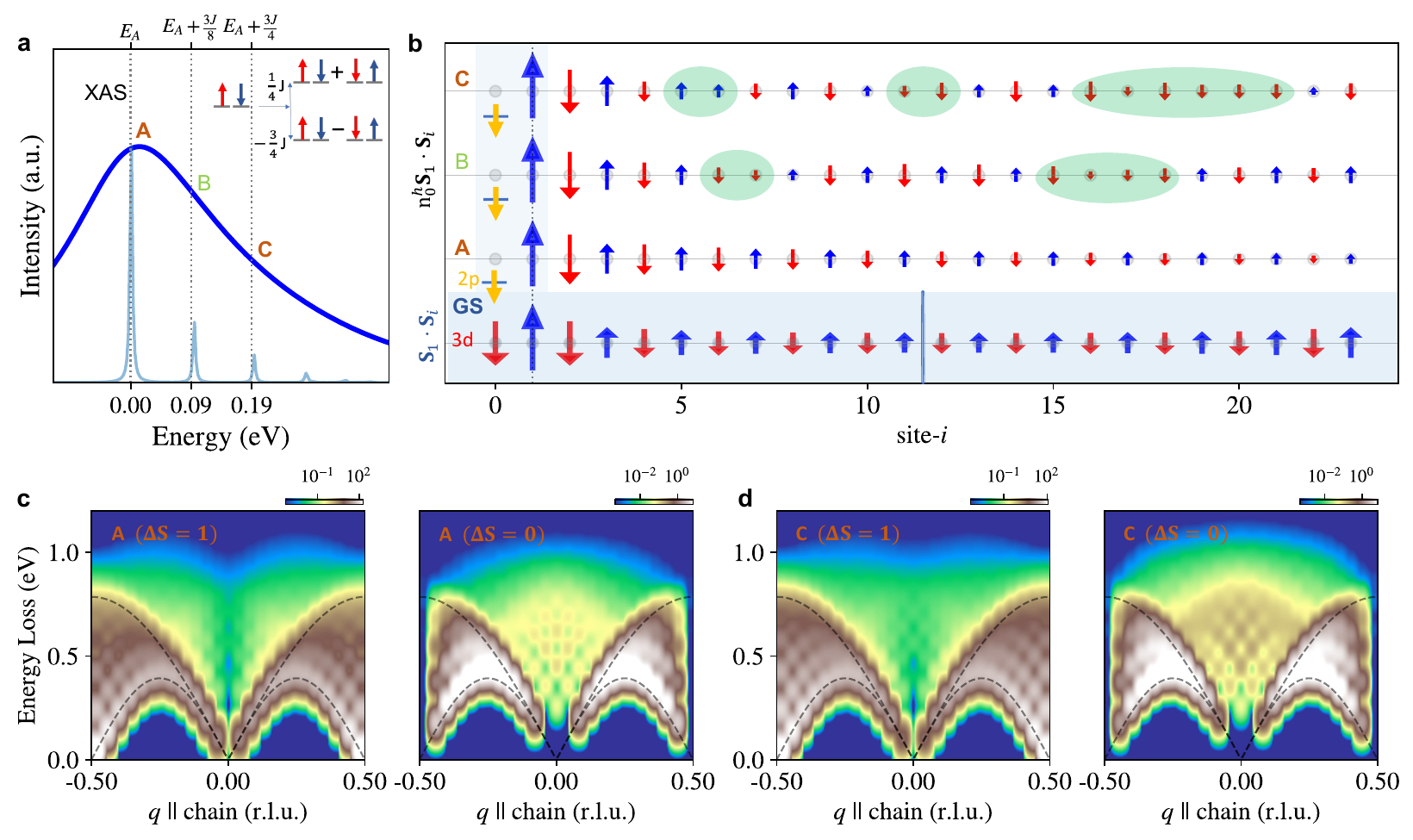}

    \caption{ ({\bf a}) Calculated \gls*{XAS} spectra for the half-filled $t$–$J$ model (solid blue). The total spectra has contributions from three distinct final states labeled A, B, and C, which can be seen by artificially reducing the lifetime broadening (light blue). ({\bf b}) core-hole-spin–spin correlations [see Eq.~\eqref{eq:ch-spin-spin_correlations}] 
    for the three states marked in panel \textbf{a}. The results are plotted as a function of distance $i$ from the core-hole site ($i= 0$) and a spin-up hole at site $1$, as indicated by the blue up arrow in all the cases. The shaded bubbles in the top two curves highlight domain walls (spin excitations from the AFM background) that are present in the different intermediate states. Here the magnitude of the spins have been rescaled by $1/\sqrt{|S_i|}$ for better visibility. ({\bf c}), ({\bf d}) the spin-resolved \gls*{RIXS} spectra taken for incident energies tuned to the A ($\Delta\omega = 0$) and C ($\Delta \omega > 0$)
    resonances, respectively. Results are shown in both the NSC and SC channels, respectively. The lines indicate the two-spinon continuum boundaries.}
\label{fig:fig_theory}
\end{figure*}

To understand the microscopic origin of the distinct resonant energies and detuning energy profiles, we
modeled the \gls*{RIXS} spectra with a \gls*{1D} $t$-$J$ model, which has been used previously to describe \gls*{RIXS} experiments on \SCO~\cite{Schlappa_NatComm_2018, Umesh_NewPhy_2018}. The details of this model are provided in the Methods section,  Sec.~\ref{method:rixs_modeling}. 

Figure~\ref{fig:fig_theory} shows how varying the incident energy selects different intermediate states and magnetic excitations in the $t$–$J$ chain. The calculated \gls*{XAS} spectrum, shown in Fig~\ref{fig:fig_theory}\textbf{a}, is notably asymmetric, with contributions from a dominant resonance peak~A followed by weaker features at higher incident  energy labeled  B and C. Note, the calculated \gls*{XAS} spectrum has been shifted to place peak A at zero incident energy as the location of this peak is arbitrarily set by the energy of on-site energy of the core-hole at our level of modeling. Peak A corresponds to the well-known upper Hubbard band resonance and its location $ E_\text{A}$ is determined by the energy required to excite the core electron into the upper Hubbard band creating a local $d^{10}$ configuration under the influence of the core-hole potential. The higher-energy features are found near $E_\mathrm{B} \approx E_\mathrm{A} + \frac{3}{8}J$ and $ E_\mathrm{C} \approx E_\mathrm{A} + \frac{3}{4}J$.
These energy scales are phenomenologically reminiscent of \gls*{XAS} final states where some number of spin singlets have been excited toward parallel alignment in the chain, as shown in Fig.~\ref{fig:fig_theory}(a) inset. Heuristically, B and C can be viewed as final states where some number of spinons are excited in the system prior to the creation of the core hole, resulting in short-range ferromagnetic-like segments with an associated exchange-energy cost. In particular, the energy separation of peak C matches the characteristic energy scale associated with the energy cost of breaking a singlet, suggesting that such local configurations may qualitatively contribute to additional spin-excitation observed in the spectrum.

We can place this interpretation on firmer footing by examining the expectation value of the core-hole-spin-spin correlation function $C_i = \bra{n} {n}_0^h \mathbf{S}_1 \cdot \mathbf{S}_i\ket{n}$, as shown in Fig.~\ref{fig:fig_theory}\textbf{b}. 
The top three rows indicate the value of $C_i$ evaluated for the \gls*{XAS} final (or \gls*{RIXS} intermediate) state $\ket{n}$ associated with each observed \gls*{XAS} resonance. Here the orientation and size of the arrows indicate the sign and magnitude of $C_i$. For reference, the bottom row of Fig.~\ref{fig:fig_theory}\textbf{b}, labeled ground state, shows the spin-spin correlation function spin $\bra{g} \mathbf{S}_1 \cdot \mathbf{S}_i\ket{g}$ evaluated for the ground state of the  Heisenberg spin-1/2 chain, which exhibits the expected algebraically decaying antiferromagnetic correlations. The correlation function associated with the A resonance resemble the ground state correlations as one tracks away from the core-hole site. In contrast, the correlations for the B and C resonances have several \gls*{AFM} domain walls (parallel-spin segments) distributed throughout them. The \gls*{NSC} [$\Delta S_\text{tot }=1$] spin-flip processes that are  dominant in Cu $L_3$-edge~\cite{Ament2009theoretical} will predominantly excite spin excitations within the two-spinon continuum similar to those probed by \gls*{INS}. Thus, the intermediate states associated with the A resonance are likely to produce two-spin correlations while intermediate states associated with the B and C resonances will produce multi-spin correlations. 

\begin{figure}[t]
    \centering
      \includegraphics[width=\linewidth]{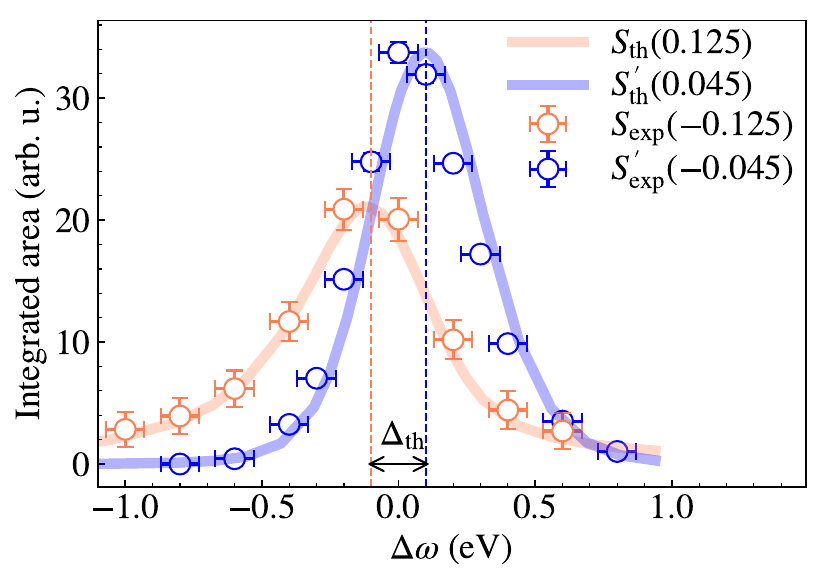}
    \caption{Experimentally measured $S_\text{exp}$ and $S^{\prime}_\text{exp}$ resonant energy profiles (circles) and theoretically calculated $S_\text{th}$, and $S^{\prime}_\text{th}$ (solid lines) at representative momentum transfers $q = -0.045$ and  $q = -0.125$ r.l.u.. The experimental data shown here are the same as in Fig.~\ref{fig:fig2}\textbf{g} and \ref{fig:fig2}\textbf{h}.  
    The predicted difference in the resonance energies $\Delta_\text{th}$ and the different lineshapes observed in the experiment are fully reproduced by the $t$-$J$ model, using previously agreed upon parameters~\cite{Schlappa_NatComm_2018} without further tuning.}
    \label{fig:fig4}
\end{figure}

To confirm this, Fig. \ref{fig:fig_theory}\textbf{c} and ~\ref{fig:fig_theory}\textbf{d} show the computed \gls*{RIXS} spectra at the A and C resonance conditions, respectively, where we have isolated the contributions from the \gls*{NSC} [$\Delta S_\text{tot}=1$] and \gls*{SC} [$\Delta S_\text{tot}=0$] channels. Note, we have plotted the intensities on a log scale to highlight the multi-spinon excitations. The two resonances produce markedly different momentum-resolved responses. In particular, spectral weight near $q\simeq 0$ is substantially enhanced at the C resonance compared to A, underscoring the role of resonance C in amplifying the long-wavelength nonlinear magnetic response. Figure~\ref{fig:fig4} directly compares the detuning behavior predicted by the $t$-$J$ model with the experimental data at $q = -0.125$, and $-0.045$ r.l.u.. The $S_\text{exp}$ and $S^{\prime}_\text{exp}$ at $-0.125$ and $-0.045$  are the same datasets as shown in Fig. \ref{fig:fig2}\textbf{g} and \ref{fig:fig2}\textbf{h}, while $S_\text{th}(q) = \int S^\text{NSC}(q,\omega)d\omega$  and $S^{\prime}_\text{th}(q) = \int S^\text{SC}(q,\omega)d\omega$  are the calculated $\Delta\omega$ dependent integrated \gls*{NSC} and \gls*{SC} \gls*{RIXS} responses, respectively. We use this formalism because the weight at the zone boundary is dominated by the NSC channel, whereas the zone center is dominated by the SC channel, as can be seen in Fig.~\ref{fig:fig_theory}\textbf{c}. We define the calculated resonant energy difference $\Delta_\text{th}$ as difference of the mean values of the $S_\text{th} (q=0.125)$ and $S^{\prime}_\text{th}(q=0.045)$ lineshapes. The calculated $\Delta_\text{th}$ and resonant energy profiles agree well at all measured momentum transfers, as shown for $q = 0.125$ and $0.045$ in Fig.~\ref{fig:fig4}. 

\begin{figure}[h]
    \centering
    \includegraphics[width=\linewidth]{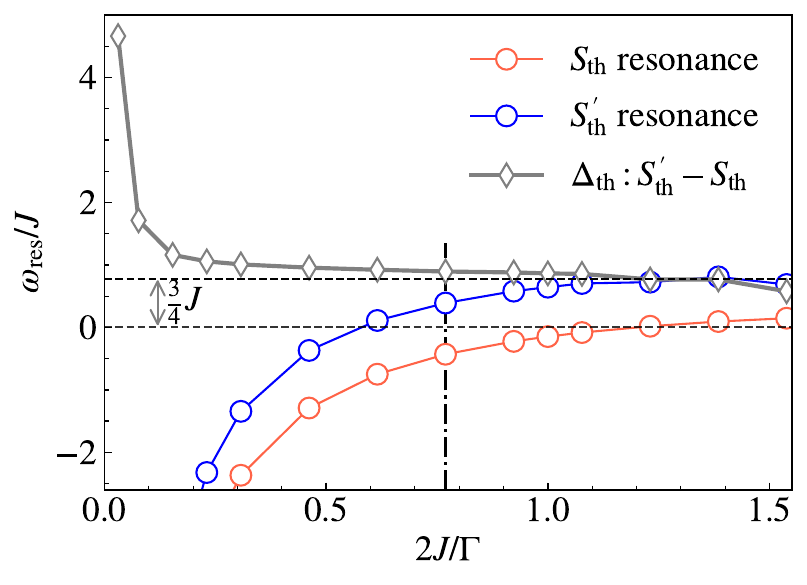}
    \caption{Calculated resonance energies $\omega_\text{res}$ of $S_\text{th}$(blue) and $S^{\prime}_\text{th}$(coral) as a function of $\frac{2J}{\Gamma}$, $\Delta_\text{th}$ (gray) represents the difference of their resonance energy.  The vertical dashed line correspond to \SCO value $\frac{2J}{\Gamma} \approx 0.8$ and the dotted horizontal lines highlights a region for which $\Delta_\text{th} = \frac34 J$ scales with $J$.}
    \label{fig:fig5}
\end{figure}

Figure~\ref{fig:fig5} demonstrates that $\Delta_\text{th} \approx 0.19$ eV ($\approx \tfrac{3}{4} J$ for \SCO) and depends primarily on $J$ over range of values when $\Gamma/2$ is fixed to a value appropriate for the Cu-$L_3$ edge. An important aspect here is that $\Delta_\text{th}$ diverges for $\frac{2J}{\Gamma} \leq 0.3$. As such, for materials with $J \ll \Gamma/2$, $\Delta_\text{th}$ does not scale with $J$, and distinct resonant energies could be resolved in those cases.  The \gls*{FWHM} for the $S_\text{th}$ also depends on $\Gamma/2$; however, it scales linearly with $J$ 
for $S^{\prime}_\text{th}$ in the regime of our interest. We show the dependence of resonant energy profiles \gls*{FWHM} on $J$ and $\Gamma/2$ for both $S_\text{th}$ and $S_\text{th}^{\prime}$ in Fig.~\ref{fig:fwhmanalysis}.

These dependencies can be rationalized as follows. The two-spin dynamical response is mediated by the core-hole annihilation via the usual spin-flip channel (enabled by the strong spin-orbit coupling in the $2p$ core level~\cite{Ament2009theoretical}). These excitations, therefore, are largely insensitive to the intermediate state lifetime $\Gamma/2$. In contrast, the higher-order multi-spin dynamical correlations can be enhanced in \gls*{RIXS} due to the intermediate state effect specific to the resonant nature of the probe~\cite{Schlappa_NatComm_2018, Umesh_NewPhy_2018, Umesh_PRB_2022}. The intermediate state lifetime is determined by $\Gamma/2$ and magnetic correlations are determined by $J$. Therefore, the competition between timescales set by $J$ and $\Gamma/2$ influence multi-spin channel creation. Crucially, because $\Gamma/2$ is an electronic structure property of an atomic species, it remains largely unaffected by the solid state effects and independent of the host material. Consequently, the incident energy detuning methodology applied to  Heisenberg spin-1/2 chain enables us to precisely disentangle the multi-spin dynamic correlations in the entire phase space of spin excitations, as established by its unique dependence on $J$ and $\Gamma/2$.

\section{Implication on fractionalization and universal quantum dynamics}

Fractional spin excitations are considered to be hallmark of spin liquid ground states and are best understood in  Heisenberg spin-1/2 chain~\cite{Michael_PRB_1997, Caux_JSM_2006,Walters_NatPhys_2009,Lake_PRL_2013,Mourigal_NatPhys_2013,Knolle_ARCMP_2019,Broholm_Science_2020}. While two-spin correlations measured in \gls*{INS} provide circumstantial evidence of characteristic continuum of magnetic excitations~\cite{Knolle_ARCMP_2019},  \gls*{INS} cannot distinguish different (two- and four-spinon) fractional excitations or broadened excitation spectra that arise from extrinsic factors like disorder that can be difficult to distinguish from a continuum.  One can only infer that the measured continuum of magnetic excitations is consistent with the magnetic sum rules predictions of the Bethe ansatz~\cite{Mourigal_NatPhys_2013}. Directly resolving the multi-spinon excitations within the two-spinon continuum would present a significant experimental leap. It is this critical aspect that is addressed in this study. 


Due to the non-linear response measured \gls*{RIXS}, the  magnetic sum rules for $S(\boldsymbol{q},\omega)$ do not apply to it. Nevertheless, the dominant low-energy excitations it produces closely resemble the spinon continuum measured in \gls*{INS}, as shown in Fig.~\ref{fig:fig1}\textbf{c}. We show that the spectral weight distribution within the continuum is tunable through selecting off-edge incident photon energies. This effect is clearly demonstrated at the \gls*{BZ} boundary, where the  continuum extends across the largest energy range, and also reproduced at intermediate momentum transfer $q=-0.125$. The spectral weight close to the lower boundary of the spinon continuum resonates at the calculated two-spin correlation resonant energy and the resonant energy profile also matches with the calculated profile for the two-spin correlation expectations, see Fig.~\ref{fig:fig4}. However, both resonant energy and resonant profile of the spectral weight at higher energy, close to the upper boundary of the spinon continuum, follows the nonlinear response corresponding to multi-spin correlations present at the  \gls*{BZ} center, both agreeing with the calculated resonant energy and profile. This behavior suggests that \gls*{RIXS} unveils two types of excitations, arising from two-spin and multi-spin correlations describing the linear and non-linear response of  Heisenberg spin-1/2 chain, respectively. As highlighted in Fig.~\ref{fig:fig2}\textbf{e-f}, these are fully disentangled inside the two-spinon continuum. The differing resonant energy and resonant energy profiles deduced from magnetic excitation spectrum can also help bound methods of calculating correlation function.

Moreover, it has been recently suggested that higher-order many-body correlations are essential to discover universal scaling behavior of  Heisenberg spin-1/2 chains~\cite{Nardis_PRL_2023, Rosenberg2024, Krajnik_PRL_2024} and an important observable to experimentally characterize  quantum many-body systems~\cite{Bohnet_Science_2016, Schweigler_Nature_2017}. The multi-spin dynamical correlations accessed by \gls*{RIXS} are related to higher-order spin correlations of type $S_i^z[\mathbf{S}_i\cdot (\mathbf{S}_{i+1}+\mathbf{S}_{i-1})]^k$ and $[\mathbf{S}_i\cdot (\mathbf{S}_{i+1}+\mathbf{S}_{i-1})]^k$ in the \gls*{NSC} and \gls*{SC} channels, respectively, where $k$ is some power~\cite{Umesh_PRB_2022}. In this regard, our work provides a methodology for isolating and interrogating prevalent higher-order many-body correlations in solid-state strongly correlated systems.  

\section{Conclusion and Outlook}

Using Cu $L_3$-edge \gls*{RIXS} on Sr$_2$CuO$_3$, a  prototypical  Heisenberg spin-1/2 chain compound, we demonstrate incident-energy detuning as an experimental knob to selectively enhance or suppress two- and multi-spin components of the magnetic response. By tracking the detuning dependence of the magnetic continuum, we identify two complementary fingerprints that separate predominantly two-spin and multi-spin magnetic excitations: (i) a systematic shift of the resonance condition between the two components, and (ii) qualitatively different resonance profiles in incident energy. In particular, the two-spin dominated response exhibits a Lorentzian-like detuning dependence, whereas the additional multi-spin spectral weight has a Gaussian-like profile, the latter arising from distinct intermediate-states that contain magnetic excitations. A key result is that the resonance conditions for the two- and multi-spin contributions are separated by an energy scale of order the magnetic exchange. For materials with weak (super)exchange interactions ($J \ll \Gamma/2$), the resonance conditions do not scale down with $J$, allowing distinct resonant energies to be resolved experimentally. The measured separation is consistent with an energy cost $\approx \frac{3}{4}J$, naturally interpreted as the cost of breaking a local singlet in the antiferromagnetic background. This interpretation is corroborated by the distinct intermediate-state spin correlations accessed at the corresponding resonances and is quantitatively reproduced by exact diagonalization of the one-dimensional $t$--$J$ model. Taken together, these results demonstrate that detuning can be used not only to reveal multi-spin dynamics outside the two-spinon continuum, but also to expose the contributions from these excitations \textit{within} the continuum by tuning the relative weight of linear (two-spin) and nonlinear (multi-spin) channels.

Looking forward, the detuning-based approach introduced here is broadly applicable to other strongly correlated materials, leveraging  \gls*{RIXS} sensitivity to intermediate-state selection rules and competing timescales. In particular, extending this methodology to higher-dimensional systems could provide a practical route to isolating multi-particle correlations in thin films and transient states relevant to candidate Kitaev materials and other fractionalized phases~\cite{Halasz_PRB_2016, Natori_PRB_2017}. The same control knob is also promising for time-resolved \gls*{RIXS} at X-ray free-electron lasers: by choosing incident energies that strongly suppress the multi-spin channel, one can simplify the interpretation of pump-induced changes by emphasizing the two- or multi-particle responses. More generally, tunable access to multi-spin correlations creates opportunities to connect \gls*{RIXS} measurements to quantum-information diagnostics (e.g., entanglement witnesses and quantum Fisher information)~\cite{ren2024witnessing} and motivates strategies to engineer the intermediate state---for example by modifying the effective core-hole lifetime in cavity-based geometries~\cite{Huang_PRR_2021}. Finally, the intermediate-state perspective developed here should generalize beyond the Cu $L_3$-edge, with natural extensions to Cu $K$-edge and O $K$-edge \gls*{RIXS} in transition-metal oxides.

\section{Methods}

\subsection{Sample details}
\SCO \, crystallizes into an orthorhombic I\textit{mmm} space group, with lattice parameters of \textit{a} = 3.9089 \AA, \textit{b} = 3.4940 \AA, and \textit{c} = 12.6910 \AA~\cite{Ami_PRB_1995}. The spin-1/2 chains are formed by the Cu$^{2+}$ ions running along the crystallographic \textit{b} direction. The exchange interaction between the spins within the chain is estimated to be $J = 241(11)$~meV~\cite{Walters_NatPhys_2009}.  Small residual inter-chain interactions drive the system to a three-dimensional N{\'e}el ordering at $T_\text{N} = 5.4$~K~\cite{Kojima_PRL_1997}. In this work, all the measurements are carried out at $T=35$ K, where the system is well described by the isotropic Heisenberg spin-1/2 chain model~\cite{Walters_NatPhys_2009,Schlappa_Nature_2012}.  




\subsection{X-ray Absorption (XAS) and resonant inelastic x-ray Scattering (RIXS)}
High-resolution \gls*{XAS} and \gls*{RIXS} experiments were carried out at the SIX 2-ID beamline of NSLS-II. The experimental energy resolution at the Cu $L_3$-edge  was 33 meV for the \gls*{RIXS} measurements, determined by the full width at half-maximum of the elastic peak measured from a reference multilayer sample. The error on incident energy ($\sigma_\omega{(\text{incident})}$) is $\approx$ 50 meV. All \gls*{RIXS} spectra shown in this study have been corrected for self-absorption, following previously established procedure~\cite{Bhartiya2025}. \\

\subsection{RIXS cross-section modeling}\label{method:rixs_modeling}
We model the system using the half-filled $t-J$ model. Its Hamiltonian, written in hole language~\cite{Schlappa_NatComm_2018, Umesh_NewPhy_2018}, is given by
\begin{equation}\label{eq:tJHamiltonian}
	H=-t\sum_{\langle i,j\rangle,\sigma} d^\dagger_{i,\sigma}d^{\phantom\dagger}_{j,\sigma}+ J\sum\limits_{i}\big(\boldsymbol{S}_i\cdot \boldsymbol{S}_{i+1} - \frac{1}{4} n_i n_{i+1}\big).  
\end{equation}
Here, $d^\dagger_{i,\sigma}~(d^{\phantom\dagger}_{i,\sigma})$ is the creation (annihilation) operator for a spin-$\sigma$ ($= \uparrow,\downarrow$) hole at site $i$ under the constraint of no double occupancy; $t$ and $J$ are the nearest-neighbor hopping integral and exchange coupling, respectively; $\langle i, j\rangle$ denotes a sum over nearest neighbors; $n_{i}=\sum_{\sigma}d^\dagger_{i,\sigma}d^{\phantom\dagger}_{j,\sigma}$ is the number operator; and $\boldsymbol{S}_{i}$ is the spin operator at site $i$. Throughout, we use $J=0.25$~eV to reproduce the energy-momentum distribution of the two-spinon continuum in our \gls*{RIXS} experiments and fix $t=0.3$ eV, which is typical for cuprates. When calculating the intermediate state of the \gls*{RIXS} process we supplement Eq.~\eqref{eq:tJHamiltonian} to include the interaction between the core and valence holes via an effective Coulomb interaction  
\begin{equation}\label{eq:Hintermediate}
    H_c = H + V_\mathrm{ch} \sum_{i,\sigma} n_{i\sigma} n_{i\sigma}^p+  \epsilon_\text{ch} \sum_{i,\sigma} n_{i\sigma}^p, 
\end{equation}
where $V_\mathrm{ch}=6.67t$ characterizes its overall strength and $\epsilon_\text{ch}$ accounts for the local transition energy between $2p_{3/2}\rightarrow 3d$ at Cu $L_3$-edge and can be accounted for by an arbitrary constant shift in our formalism. Here $n_{i\sigma}$ and $n_{i\sigma}^p$ are the hole number operators for the 3d $d$ and core $2p$ orbitals,  respectively, of the Cu atom. Throughout, we solved Eqs.~\eqref{eq:tJHamiltonian} and \eqref{eq:Hintermediate} on $L = 24$ site chain using \gls*{ED}. We evaluated the \gls*{RIXS} cross-section  using the Kramers-Heisenberg formalism~\cite{Ament_RMP_2011, Mitrano2024exploring},  where the intensity is given by 
\begin{equation}\label{eq:KH}
I(q,\Omega, \omega_\text{in}) \propto \sum\limits_{f}\left\lvert
M_{f,g}\right\rvert^{2}
\times \delta\left(E_{f} - E_{g} - \Omega\right).   
\end{equation}
where 
\begin{equation}
    M_{f,g}=\sum\limits_{m}\frac{\langle f \lvert D^{\dagger}_{k_\mathrm{out}}\rvert m \rangle \langle m \lvert D^{\phantom\dagger}_{k_\mathrm{in}} \rvert g \rangle}
{\omega_\mathrm{in}  - (E_{m} - E_{g}) + \mathrm{i} \Gamma_{m}/2 }
\end{equation}
is the scattering cross-section. 
Here, the incoming (outgoing) photons have energy $\omega_\mathrm{in} =\omega_\text{res}+\Delta\omega$ ($\omega_\mathrm{out}$) and momentum $k_\mathrm{in}$ ($k_\mathrm{out}$). 
$\Omega = \omega_\mathrm{in} - \omega_\mathrm{out}$ and 
$q = k_\mathrm{in}-k_\mathrm{out}$ are the energy and momentum transferred along the chain direction, respectively. $\lvert g \rangle$, $\lvert m\rangle$, and $\lvert f \rangle$ are the initial, intermediate, and final states of the \gls*{RIXS} process with energies $E_{g}$, $E_m$, and $E_{f}$, respectively. $D_{k}= \sum_{i, \sigma} e^{\mathrm{i} k r_i} D_{i,\sigma}$ is the dipole operator describing the $2p\rightarrow 3d$ transition with $D_{i,\sigma} = \sum_{\alpha} d^{\phantom\dagger}_{i \sigma} p_{i,\alpha}^\dagger$, where $p^\dagger_{i,\alpha}$ is the creation operator on the $2p_{i, \alpha}\in L_3$ core level. $\Gamma_m/2$ is the core-hole broadening of intermediate state $m$ and related to the inverse core-hole lifetime. For our numerical calculations we assume $\Gamma_m/2 = 0.3$ eV for all $m$, which is typical for calculations at the Cu $L_3$-edge. We neglect the mobility of the core hole in the intermediate state. 

In the limit $\Gamma_{m} \rightarrow 0$ and $\Delta_m  \equiv E_m - E_g \rightarrow \omega_\text{in}$,  the action of the dipole operator can be understood as a local quench and energy resolved, where $\ket{m, i }$ can be considered as the effective ground state, and the \gls*{RIXS} spectra can be rewritten as 
\begin{equation}
I(q, \omega) = \sum_f |\langle f| \sum_{m, i} e^{\text{i} q r_i} d_{i}^\dagger p^{\phantom\dagger}_i |m \rangle  |^2 \Lambda_{m,i}^2(\omega_\text{in})  \delta (E_f - E_g- \omega).
\end{equation}
Here, $\Lambda_{m, i} (\omega_\text{in}) = |\langle m| d^{\phantom\dagger}_i p_i^\dagger |g\rangle |^2 \delta(\Delta_m - \omega_\text{in})$, which will be a scaling factor for a particular intermediate state $m$. Alternatively, these are the cross-sectional overlap in \gls*{XAS} shown in Fig.~\ref{fig:fig_theory}\textbf{a} and are identified as the ``A", ``B", and ``C" peaks in the \gls*{XAS}. 
Different many-body intermediate states can be targeted by selecting specific incident energies.  Notice that these core-hole sites have translation symmetry and therefore, will have degenerate spectrum. In our analysis, the intermediates state eigen spectrum involves three major components $m \in \text{A, B, C}$.  Importantly, these components are separated by energies dictated by the strength of magnetic super-exchange.

To assess the nature of the different intermediate states that are active in the \gls*{RIXS} process, we computed a core-hole-spin-spin correlation function of the form 
\begin{equation}\label{eq:ch-spin-spin_correlations}
 C_{i,0}^{m} = \langle m, i = 0 | n_0^h  {\bf S}_1 \cdot {\bf S}_i   | m \rangle 
\end{equation}
where $n^h_0~(= 1- d_{0,\sigma}^\dagger d_{0,\sigma}^{\phantom\dagger})$ annihilates a hole at the core-hole site $i = 0$, and $\ket{m}$ is the relevant intermediate state of the \gls*{RIXS} process. 
This correlation function measures the magnetic correlations between a spin located next to the core hole and one further down the chain at site $i$. These correlation function preserves the spin SU(2) symmetry.  \\



\noindent\textbf{Acknowledgments}: We thank A.~Zheludev (ETH Z{\"u}rich) for providing single crystal samples of \SCO. Work at Brookhaven National Laboratory (experiment planning and data acquisition, data analysis, and writing) was supported by the U.S. Department of Energy (DOE), Division of Materials Science, under Contract No. DE-SC0012704. Work at the University of Tennessee (theory modeling, data interpretation, and writing) was supported by the U.S. Department of Energy, Office of Science, Office of Basic Energy Sciences, under Award No. DE-SC0022311. This research used beamline 2-ID of the National Synchrotron Light Source II, a U.S. DOE Office of Science User Facility operated for the DOE Office of Science by Brookhaven National Laboratory under Contract No. DE-SC0012704.  M. M. was supported by the US Department of Energy, Office of Basic Energy Sciences, Early Career Award Program, under award no. DE-SC0022883. U.~K. and S.~O. were supported by the U.S. Department of Energy, Office of Science, National Quantum Information Sciences Research Centers, Quantum Science Center.\\

\noindent\textbf{Author Contributions}: The research project was conceived by V.~K.~B., U.~K., S.~J., and V.~B.. V.~K.~B., T.~K., S.~F., J.~P., and V.~B. performed the \gls*{RIXS} measurements. V.~K.~B. analyzed the \gls*{RIXS} data with guidance from J.~P., M.~P.~M.~D., I.~Z., and V.~B.; U.~K. and S.~J. performed the theory calculations, and worked on the data interpretation together with V.~K.~B. and V.~B.; V.~K.~B., U.~K., S.~J., and V.~B. wrote the manuscript with input from all authors. S.~J. and V.~B. supervised the theory and experimental components of the work, respectively. 


\bibliography{VK_library}

\appendix
\renewcommand\thefigure{\thesection.\arabic{figure}}    
\counterwithin{figure}{section}

\section{Alternative fitting of resonant energy profiles}
\label{sec:fitting_detuning}

\begin{figure}
    \centering
    \includegraphics[width=\linewidth]{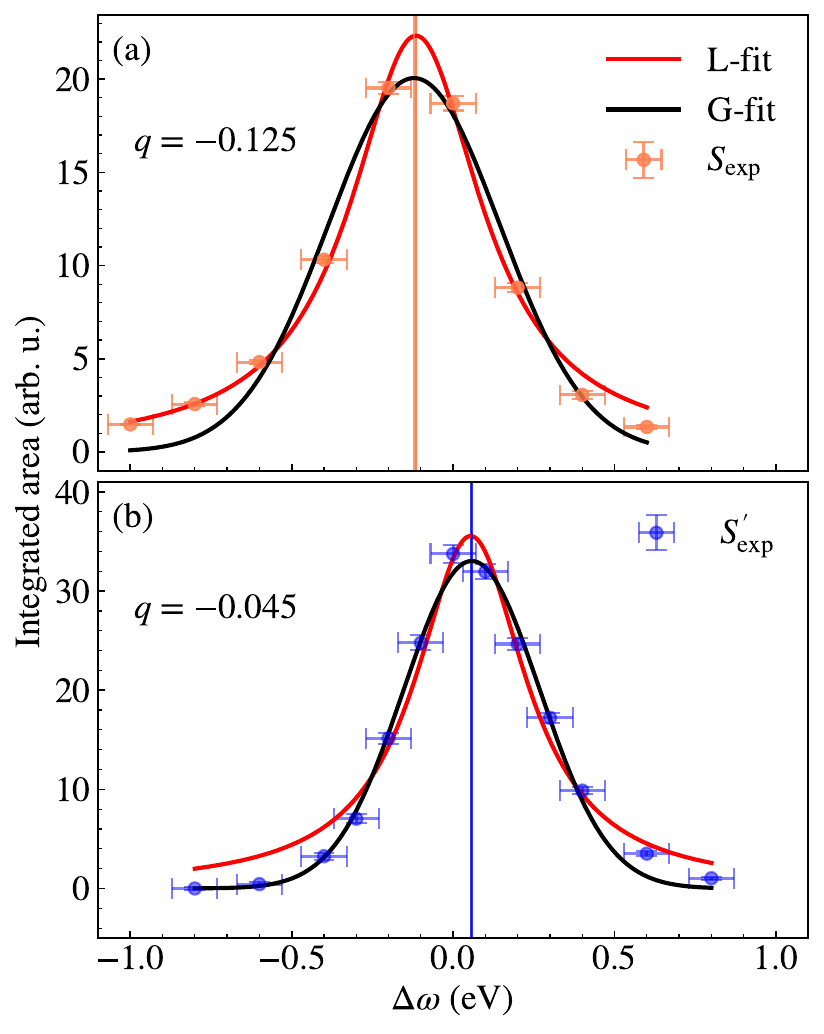}
    \caption{Experimentally measured resonant energy profiles, (a) $S^{\prime}_\text{exp}$ at $q = -0.125$ r.l.u. and (b) $S^{\prime}_\text{exp}$ at $q = -0.045$ r.l.u.. These datasets are the same as in Fig.~\ref{fig:fig2}\textbf{g}. Both resonant energy profiles are fitted to Lorentzian (red) and Gaussian (black) lineshapes, labeled as  L- and G- fit, respectively.}    \label{fig:fitting_detuning}
\end{figure}

In Fig.~\ref{fig:fitting_detuning}, we present the fitting of $S_\text{exp}$ and $S^{\prime}_\text{exp}$ resonant energy profiles  with alternative  line shapes. It is evident that the best fit can only be achieved with the Lorentzian lineshape for $S_\text{exp}$ and with the Gaussian line shape for $S^{\prime}_\text{exp}$. While both lineshapes reproduce the same mean value (resonant energy), the tails are better captured by the specific line shape.

\section{Additional details on FWHM of resonant energy profiles}
\begin{figure*}
    \centering
     \includegraphics[width=0.9\linewidth]{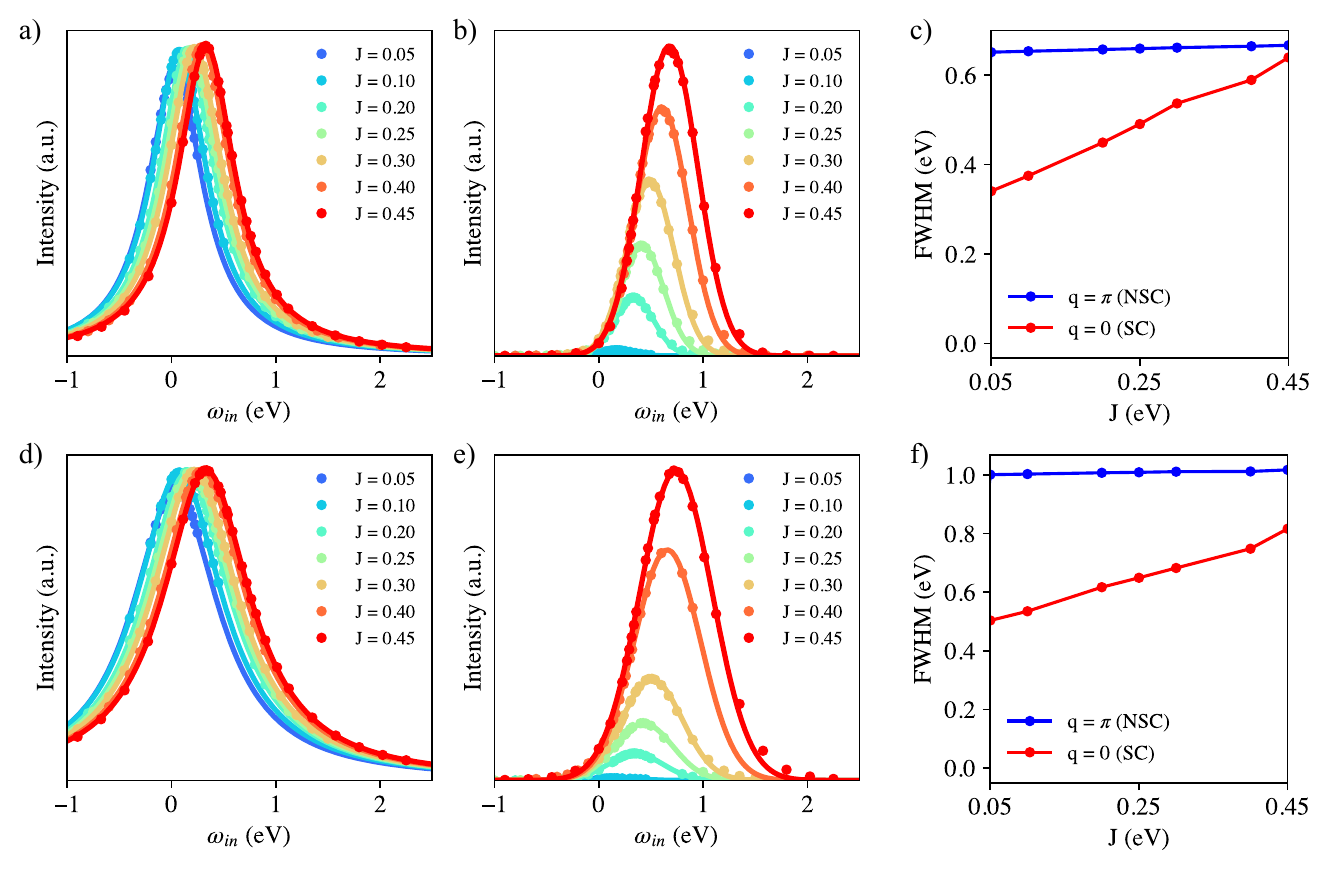}
    \caption{Broadening of resonance condition of the NSC (two-spin) and SC (multi-spin) channels evaluated using Eq.~\ref{eq:KH}.  Panels (a-c) show the data evaluated using $\Gamma/2 = 0.325$ eV and panels (d-f) for 0.5 eV. (a) and (d) show the broadening of the resonance condition in the NSC channel at $q=0.5$, whereas, (b) and (e) show the broadening of the resonance condition in the SC channel at $q= 0$. Panel (c) and (f) show the fitted FWHM for $\Gamma = 0.325$ eV and 0.5 eV, respectively. Fitting for the  NSC and SC channels is carried out using Lorentzian and Gaussian fittings, respectively.}
    \label{fig:fwhmanalysis}
\end{figure*}



Fig.~\ref{fig:fwhmanalysis} shows the calculated resonant energy profiles for the NSC and SC channels of \gls*{RIXS} cross-section at two fixed values of $\Gamma/2 = \{0.325, 0.5\}$ eV and  for a range of $J$ values. To remind, NSC channel is predominantly of two-spin character and like INS, however SC channel is of multi-spin character and \gls*{RIXS} specific. Panels (a-c) show the results for  $\Gamma/2 = 0.325$ eV. Panel (a) shows the incident energy dependence of integrated \gls*{RIXS} spectra   for the NSC channel (at $q$ = $0.5$). The resonant energy line shapes fit well with a Lorentzian function.  Similarly, panel (b) shows the results for the SC channel (at $q$ = $0$). In this case,  the resonant energy lineshapes fit well with a Gaussian function. Panel (c) shows the fitted FWHM for these NSC and SC resonant energy profiles. Notice that  FWHM = $\Gamma$ for the NSC channel is independent of  $J$. On the other hand for SC channel (at $q=0$), fitted FWHM appears to increase linearly with $J$ starting from FWHM = $\Gamma/2$ at small $J$. The results shown in panels (d-f) are for $\Gamma/2 =0.5$ eV and show a  consistent behavior with the $\Gamma/2 =0.5$ eV case. Therefore validating the outcome that FWHM of two-spin channel depends only on core-hole lifetime $\Gamma/2$ and for multi-spin channel on both exchange interaction and core-hole lifetime $\sim \frac{J}{\Gamma/2}$.



\section{Two-spin dynamical structure factor}\label{sec:DSF}

The magnetic excitation spectrum of  Heisenberg spin-1/2 chain, within the linear response,  can be described by the fractional spin excitations called spinon and its  dynamical structure factor $S(q,\omega)$~\cite{Michael_PRB_1997,Caux_JSM_2006} is given by
\begin{equation}
S(q,\omega)=M(q,\omega,J) D(q,\omega,J) .
\end{equation}
Here $q$ is the momentum transfer along the chain and $J$ is the exchange coupling constant. $M(q,\omega,J)$ is the momentum-frequency dependent spectral weight  and  $D(q,\omega)$ the density of states, calculated within the linear response of  Heisenberg spin-1/2 chain. $M(q, \omega, J)$ is given as
\begin{equation}\label{eq:Mqw}
M(q, \omega, J) = 0.5 \exp(-I(t(q, \omega, J)))
\end{equation}
where the parameter $t(q, \omega, J)$ is defined as
\begin{widetext}
\begin{equation}
t(q, \omega, J) = 
\begin{cases}
\frac{4}{\pi} \operatorname{arccosh}\left( \sqrt{\frac{\omega_\mathrm{U}(q, J)^2 - \omega_\mathrm{L}(q, J)^2}{\omega^2 - \omega_\mathrm{L}(q, J)^2}} \right) & \text{if } \omega_\mathrm{L}(q, J) \leq \omega \leq \omega_\mathrm{U}(q, J) \\
0 & \text{otherwise}.
\end{cases}
\end{equation}
\end{widetext}

The parameters $\omega_L$ and $\omega_U$ represent the lower and upper limits of the spinon continuum, respectively, and are defined as, 
\begin{align}
\omega_\mathrm{L}(q, J) &= \frac{\pi}{2} J |\sin(2\pi q)| \\
\omega_\mathrm{U}(q, J) &= \pi J |\sin(\pi q)|
\end{align}

The integral $I(t)$ in Eq.~\ref{eq:Mqw} is given  as
\begin{equation} \label{eq:integral}
I(t) = \int_{0}^{\infty} \frac{\cosh(2x) \cos(xt) - 1}{x \sinh(2x) \cosh(x)} e^x dx
\end{equation}
and evaluated numerically using ``quad" function from ``scipy". To avoid singularity, the lower and upper integration limit are replaced with 10$^{-6}$ and 200, respectively. 

The density of states $D(q,\omega,J)$ is given as

\begin{equation}
D(q, \omega, J) =
\begin{cases}
\frac{1}{\sqrt{\omega_\mathrm{U}(q, J)^2 - \omega^2}} & \text{if } \omega_\mathrm{L}(q, J) \leq \omega \leq \omega_\mathrm{U}(q, J) \\
0 & \text{otherwise}
\end{cases}
\end{equation}


The experimentally observed lineshape is calculated by convoluting $S(q, \omega)$ with a Gaussian ($\mathcal{G}$) line shape broadened by instrument resolution

\begin{equation}
S^\text{exp}(q, \omega) = S(q, \omega) \circledast \mathcal{G}.
\end{equation}


\end{document}